\documentclass[a4paper,12pt]{article}
\usepackage{graphicx}
\usepackage{amssymb}
\begin{document}
%
%\begin{frontmatter}
\begin{titlepage}

\begin{flushright}
       LYCEN 2003-36\\
       October 15th, 2003 \\
\end{flushright}

\vfill

{\LARGE\bf \begin{center}

        Calibration of the EDELWEISS Cryogenic Heat-and-ionisation
        Germanium Detectors \\
        for Dark Matter Search 
\end{center}}

\vfill

\begin{center}
 The EDELWEISS Collaboration: \\  
 O.~Martineau$^{1}$,
 A.~Beno\^{\i}t$^{2}$,
 L.~Berg\'e$^{3}$,
 A.~Broniatowski$^{3}$,
 L.~Chabert$^{1}$,
 B.~Chambon$^{1}$,
 M.~Chapellier$^{4}$,
 G.~Chardin$^{5}$,
 P.~Charvin$^{5,6}$,
 M.~De~J\'esus$^{1}$,
 P. Di~Stefano$^{1}$,
 D.~Drain$^{1}$,
 L.~Dumoulin$^{3}$,
 J.~Gascon$^{1}$,
 G.~Gerbier$^{5}$,
 E.~Gerlic$^{1}$,
 C.~Goldbach$^{7}$,
 M.~Goyot$^{1}$,
 M.~Gros$^{5}$,
 J.P.~Hadjout$^{1}$,
 S.~Herv\'e$^{5}$,
 A.~Juillard$^{3}$,
 A.~de~Lesquen$^{5}$,
 M.~Loidl$^{5}$,
 J.~Mallet$^{5}$,
 S.~Marnieros$^{3}$,
 N.~Mirabolfathi$^{6}$,
 L.~Mosca$^{5,6}$,
 X.-F.~Navick$^{5}$,
 G.~Nollez$^{7}$,
 P.~Pari$^{4}$,
 C.~Riccio$^{5,6}$,
 V.~Sanglard$^{1}$,
 L.~Schoeffel$^{5}$,
 M.~Stern$^{1}$,
 L.~Vagneron$^{1}$
\end{center}

%
%\corauth[cor1]{Corresponding author.
% Present adress: Institut f\"ur Kernphysik, Forschungszentrum Karlsruhe,
% Postfach 3640, 76021 Karlsruhe, Germany.
% Tel.:+49-724-782-4153; fax: +49-724-782-4075; e-mail: martin@ik.fzk.de}
%

{\noindent\small
$^{1}$Institut de Physique Nucl\'eaire de Lyon-UCBL, IN2P3-CNRS,
      4 rue Enrico Fermi, 69622 Villeurbanne Cedex, France\\
$^{2}$Centre de Recherche sur les Tr\`es Basses Temp\'eratures,
      SPM-CNRS, BP 166, 38042 Grenoble, France\\
$^{3}$Centre de Spectroscopie Nucl\'eaire et de Spectroscopie de Masse,
      IN2P3-CNRS, Universit\'e Paris XI,
      bat 108, 91405 Orsay, France\\
$^{4}$CEA, Centre d'\'Etudes Nucl\'eaires de Saclay,
      DSM/DRECAM, 91191 Gif-sur-Yvette Cedex, France\\
$^{5}$CEA, Centre d'\'Etudes Nucl\'eaires de Saclay,
      DSM/DAPNIA, 91191 Gif-sur-Yvette Cedex, France\\
$^{6}$Laboratoire Souterrain de Modane, CEA-CNRS, 90 rue Polset,
      73500 Modane, France\\
$^{7}$Institut d'Astrophysique de Paris, INSU-CNRS,
      98 bis Bd Arago, 75014 Paris, France
}
\vfill

%
%\begin{abstract}

\begin{center}{\large\bf Abstract}\end{center}

Several aspects of the analysis of the data obtained with the cryogenic heat-and-ionisation 
Ge detectors used by the EDELWEISS dark matter search experiment are presented. Their calibration, the determination 
of their energy threshold, fiducial volume and nuclear recoil acceptance are detailed.

%\end{abstract}
%
%\begin{keyword}
%Dark matter \sep WIMPs direct detection \sep cryogenic detectors \sep data analysis.
%\PACS
%29.40.Wk \sep % Solid-state detectors
%07.57.K  \sep % Bolometers
%95.35.+d      %Dark matter
%\end{keyword}
%\end{frontmatter}

\end{titlepage}

\section{Introduction}
The dominant ($\sim$ 90\%) component of the mass budget of the Universe may consist in Weakly Interacting Massive Particles 
(WIMPs), which could be the Lightest Supersymmetric Particles (neutralinos in most models)\footnote{
See e.g.~\cite{berg} for a review.}. WIMPs would be present at the galactic scale
as a halo of mass typically ten times larger than the visible part of the galaxy. The EDELWEISS collaboration has developped 
heat-and-ionisation Ge detectors~\cite{xfn} to measure recoils induced by elastic scattering of
galactic WIMPs on a target nucleus. Constraints on the
spin-independent WIMP-nucleon cross-section in the framework of the 
Minimal SuperSymmetric Model (MSSM) have been derived from the nuclear recoil rate measured with the EDELWEISS detectors \cite{ed2000}, \cite{ed2002}. We present 
in this paper the experimental details of these
measurements. In Section \ref{detectors}, we describe briefly the experimental setup and the method of detection of an energy deposit in the target. We then 
show the calibration procedure for the heat and ionisation signals (Section \ref{channels}), the trigger threshold determination (Section \ref{seuil}) and 
the tagging of the nuclear recoils (Section \ref{zoneneu}). 
We finally present an original method to determine the fiducial volume of the detectors (Section \ref{volfid}). 
\section{The EDELWEISS detectors}
\label{detectors}
\subsection{Experimental setup}
\begin{figure*}
\begin{center}
\begin{tabular}[h]
{c}
            \includegraphics*[width=12.1cm,height=10cm,bb=10 0 505 456]{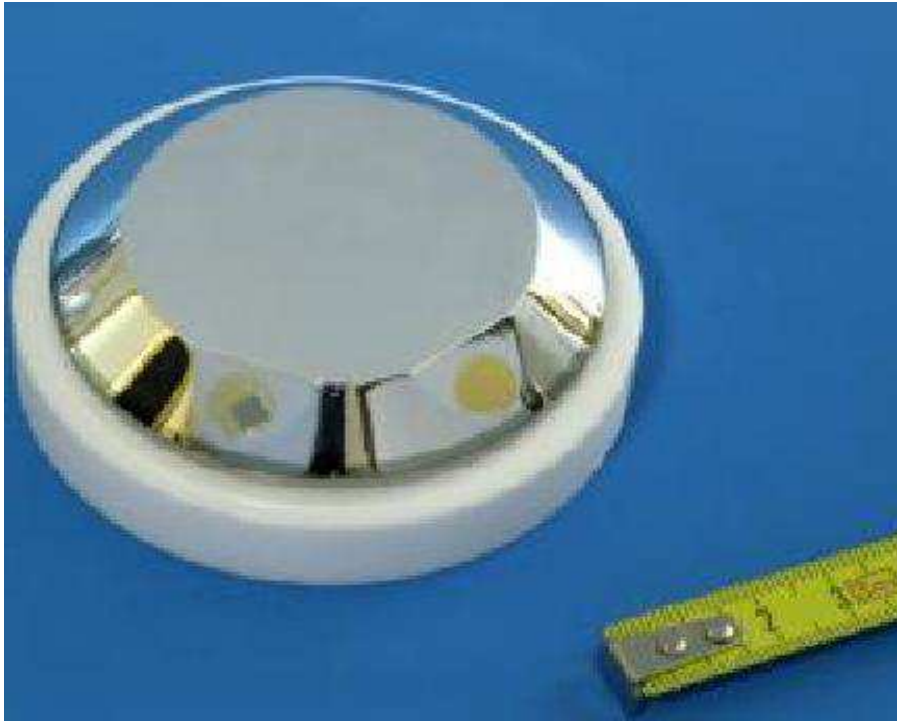} \\
            \includegraphics*[width=12.1cm,height=7.5cm,bb=5 196 570 606]{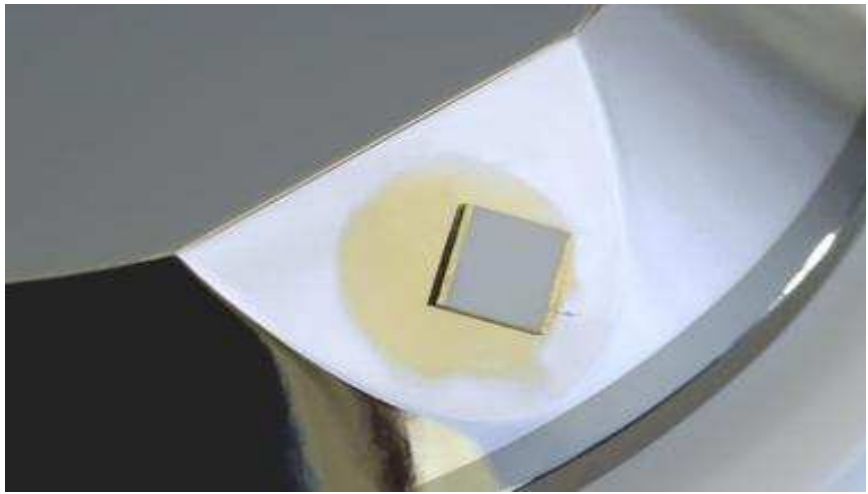}
\end{tabular}
\end{center}
\caption{
\label{detect}
\textit{
         Top pannel: EDELWEISS GGA1 detector ($\Phi=$70 mm). Bottom pannel: close-up on the NTD thermal sensor
	 glued on its golden pad on the beveled part of the crystal. 
        }
        }
\end{figure*}

The experimental setup of the EDELWEISS-I experiment is described in \cite{ed2000} and \cite{ed2002}. We simply recall that up to three 
detectors can be housed in a low background dilution cryostat working at a regulated temperature (27 mK in 
\cite{ed2000} and 17~mK in \cite{ed2002}). The EDELWEISS detectors
are made of a germanium absorber (target for the incident particles) equiped with a thermal sensor and with
metallic electrodes for charge collection. The
simultaneous measurement of both phonons and charges created by a single interaction is therefore possible. \\
The main characteristics of the detectors studied in this article are given in Table \ref{det}. For all of these detectors, the absorber is a $\sim$320~g Ge cylindrical 
crystal ($\sim$70~mm diameter and 20~mm thickness). Their edges have been beveled at an angle of 45$^o$ (Fig. \ref{detect}). 
The electrodes for ionisation measurement are made of 
100~nm Al layers sputtered on the surfaces after etching. The top electrode is divided in a central part and a guard ring, electrically
decoupled for radial localization of the charge deposition. The bottom electrode is the common reference. For the GGA1 and GGA3 detectors (GSA1 and GSA3), a 
60~nm hydrogenated amorphous germanium (silicon) layer was deposited under the electrodes in order to reduce the charge
collection problems associated with events where the energy is deposited close to the detector surface. It 
has indeed been shown that the probability that charge carriers be collected on the same-sign electrode during
the diffusion phase which preceeds the charge collection (dead layer problem) is reduced for this type of 
detectors [5, 6]. \\
The thermal sensor consists of a Neutron Transmutation Doped germanium crystal (NTD)\footnote{The NTD thermal
sensors have been produced by Torre and Mangin for GeAl6 and by Haller-Beeman Associates for the other 
detectors.}, 
close to the metal-insulator transition. It is glued on a sputtered gold pad near the edge of the bottom Al
electrode (Fig. \ref{detect}). The resistance of the DC-polarized 
GeAl6 sensor was chosen to be $\sim$ 3~M$\Omega$ for GeAl6 ($T_{running}\sim$27 mK), and ranged from 3~M$\Omega$ to 6~M$\Omega$ at 17~mK for the other detectors. 
Reliable electrical contacts and heat links have been achieved by the ultrasonic bonding of gold wires (diameter 25 $\mu$m) on 
gold pads. The thickness of these pads has been chosen to minimize the production of dislocations in the absorber caused by the bonding. 
A thermal analysis of the detectors will be published in Ref.~\cite{N03}. 
\subsection{Detection method}
The rise in temperature due to an energy deposit in the absorber gives rise to a variation $\Delta{R}$ of the 
thermal sensor resistance. When the sensor is polarized by a constant current $I$, $\Delta{R}$ then induces a 
voltage fluctuation $\Delta{V}$ across the resistor, which corresponds to the heat signal:
\begin{equation}
\label{vchal}
\Delta{V}=\Delta{R}\times{I}
\end{equation} 
The ionisation signal is obtained by collection of the electron-hole pairs created by 
the interaction in the germanium crystal polarized through a bias voltage applied to the electrodes. A low bias voltage\footnote{
During the data takings, the bias voltage applied to the top electrode varied from $\pm$~3~V to $\pm$~9~V depending on the detector.}
is required to limit the heating of the cristal due to the drift of the charge carriers, known as the Neganov-Luke effect~\cite{lneg}. \\
\ \\
The energy $E_{R}$ deposited by a particle interacting in the detector can be determined by subtracting the 
Neganov-Luke effect from the heat signal~:
\begin{equation}
\label{erec}
E_{R}=\left(1+\frac{V}{\varepsilon_{\gamma}}\right)E_{H}-\frac{V}{\varepsilon_{\gamma}}E_{I}
\end{equation}
where V is the bias voltage and $\varepsilon_{\gamma}=3$~V the mean electron-hole pair creation potential in
germanium for $\gamma$-ray interactions (electron recoils). The variables $E_H$ and $E_I$ stand respectively
for the heat and ionisation signal amplitudes calibrated for $\gamma$-ray interactions following the
procedure described in Section \ref{calib}. \\
\ \\
We define the quenching variable Q as:
\begin{equation}
\label{qdef}
Q=\frac{E_I}{E_R}
\end{equation}
This variable is of particular interest in the case of WIMP search since nuclear and electronic recoils correspond to different 
ionisation efficencies. As $E_I$ and $E_H$ are calibrated using $\gamma$-rays, $Q=1$ for electronic
interactions by definition. In the case of nuclear recoils (such as those
that would be produced by WIMP interactions), this ratio is much lower: $Q\sim0.3$. The simultaneous measurement of heat and
ionisation therefore provides an event-by-event identification of the type of recoils and thus gives an efficient method to reject the 
dominant $\gamma$-ray background. The precise definition of the rejection criteria is discussed in Section \ref{zoneneu}.
\section{Calibration and resolution of heat and ionisation signals}
\label{channels}
\subsection{Calibration of heat and ionisation channels}
\label{calib}
\begin{figure*}
%\centerline{\includegraphics*[width=12cm,height=14cm,bb=50 157 547 657]{fig/gamfig.ps}
\centerline{\includegraphics*[width=12cm,height=14cm,bb=50 157 547 657]{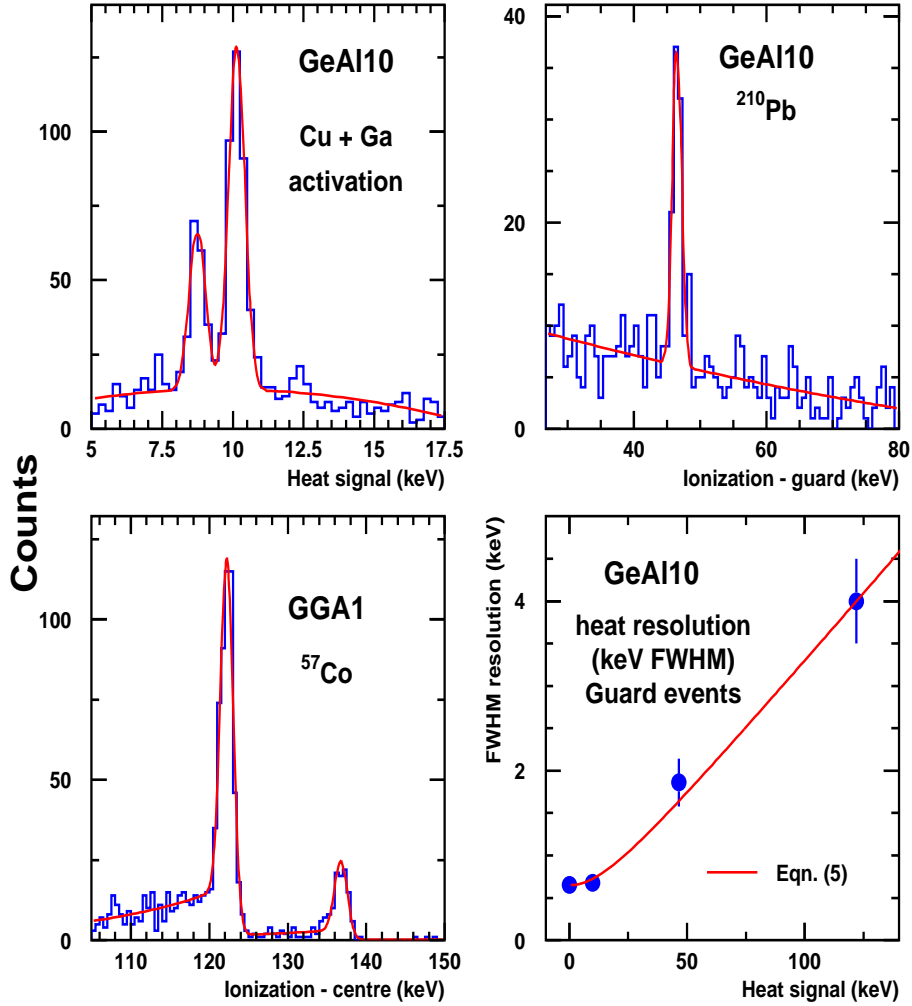}
            }
\caption{
\label{reso}
\textit{
        Figs. a), b), c): spectra obtained with the GeAl10 and GGA1 detectors in various energy regions during physics background runs (a and b) and $^{57}$Co calibration runs (c).
        For the first two spectra, the lines associated with $^{65}$Zn (8.98~keV) and $^{68}$Ge decays (10.37~keV) (a) and $^{210}$Pb contamination (46.52~keV)(b) are 
        clearly visible. In Fig. d), the baseline and peak resolutions of the GeAl10 detector heat channel for the 10.37, 46.52 and 122.1 lines are fitted 
        by the expression given in Eq. (\ref{sigt}).
           }
        }
\end{figure*}

\begin{figure*}
\centerline{
            \includegraphics*[width=8cm,height=8cm,bb=50 20 547 507]{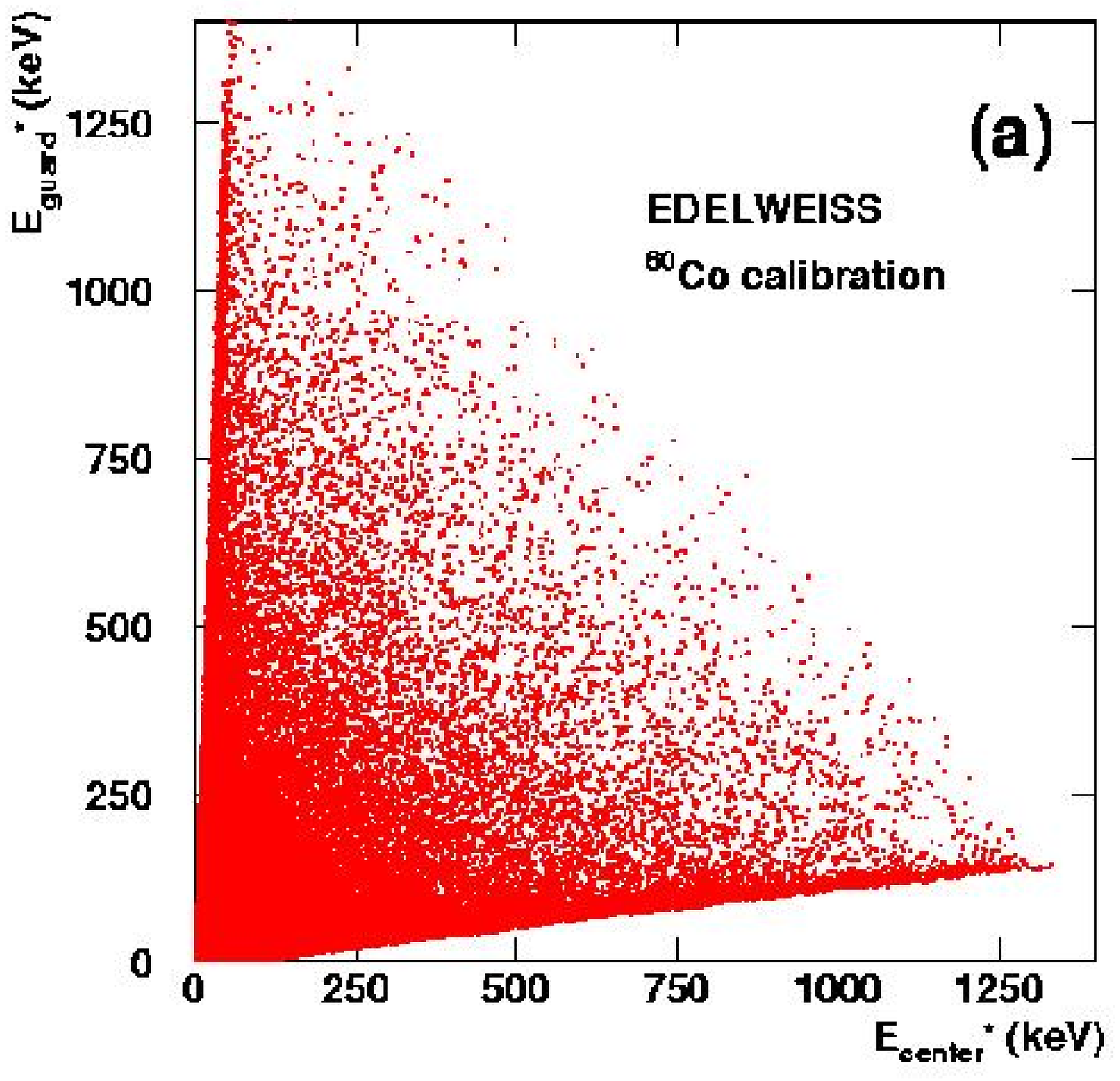}
            \includegraphics*[width=8.8cm,height=8.8cm]{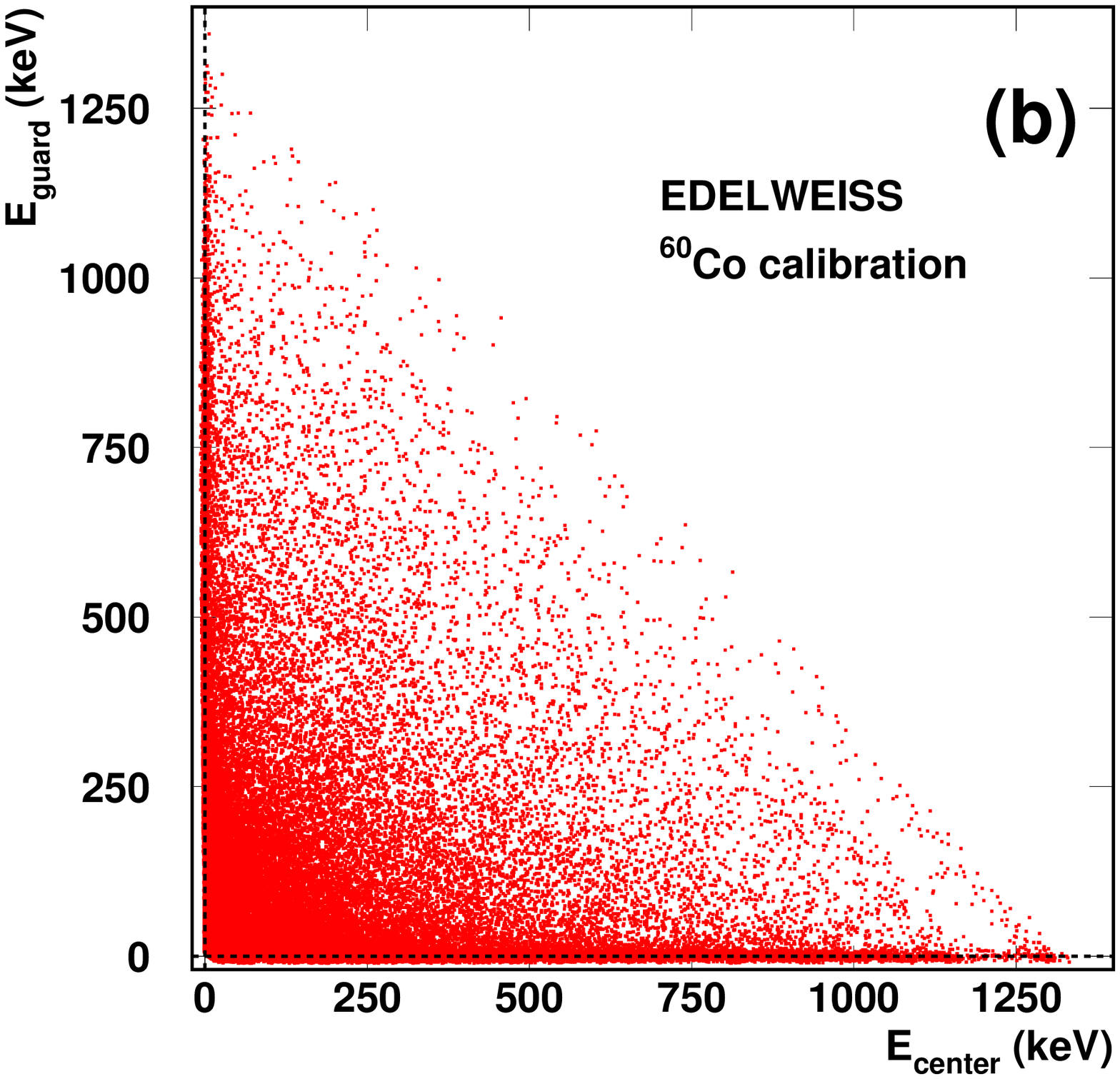}
            }
\caption{
\label{bipbrut}
\textit{
         Distribution of the guard ring versus the center electrod signals for a $^{60}$Co calibration of the GeAl6 detector at a bias
	 voltage of +6.34~V before (a) and after (b) linear correction of the cross-talk between the two channels.
         On Fig. (b), the events along the horizontal (vertical) axis correspond to center (guard) events, for
	 which the charge is fully collected on the center electrode (guard ring), 
	 and the events between the two axis correspond to shared events. Shared events represent a proportion of $\sim$ 50\% of the total number of events for
	 this calibration.
        }
        }
\end{figure*}

The ionisation signal $E_{I}$ is calibrated using a $^{57}$Co source that can be inserted in the liquid He bath of the cryostat to a distance of $\sim$ 
10~cm from the detectors, with only a $\sim$ 0.5~cm thick copper shielding layer between the source and the detectors. The 122 and 136~keV peaks 
are clearly visible on the spectra (Fig. \ref{reso}c), allowing a precise calibration of the ionisation signal. The linearity of the signal
amplitude has been verified using the 46.52~keV line from $^{210}$Pb (Fig. \ref{reso}b) in the detector
environment and the 8.98 and 10.37~keV lines from the decay of cosmic-ray induced long life isotopes 
 $^{65}$Zn and $^{68}$Ge in the detector. The calibration 
factor is observed to be stable within a fraction of percent over periods of months. 
Because of the parasite capacitance between the centre and guard electrodes, a 
charge fully collected on an electrode also induces a signal on the other. This cross-talk of a few percents 
is purely linear and remains constant in time for a given detector. 
It can thus be easily corrected off-line (Fig. \ref{bipbrut}). The ionisation signal $E_{I}$ is defined as the sum of the 
guard ring and center electrode signal amplitudes after correction of the cross-talk and calibration of the two channels. \\
The heat signal amplitude $E_{H}$ is periodically calibrated using the same $^{57}$Co source. In contrast with
ionisation, the heat signal appears to be very sensitive to long term drifts of the NTD temperature. It may for example vary by a 
few percent during several hours after transfers of cryogenic fluids. Between two $^{57}$Co  calibrations, the heat signal is therefore monitored on a continuous 
basis using the data from the low-background physics runs themselves by setting the average value of the $Q$ ratio to 1 for
electron recoils. The 46.52~keV line from $^{210}$Pb and the 8.98 and 10.37~keV lines associated with 
cosmogenesis activation of  $^{65}$Zn and $^{68}$Ge in the detector (Fig. \ref{reso}a) are used to check the quality of the
calibration of the heat signal. \\
It should be stressed again at this point that the heat and ionisation signals are calibrated using
$\gamma$-ray sources, which induce electron recoils. The $E_I$ and $E_H$ values thus correspond to the
actual energy deposit for this type of interactions only, and are therefore expressed in keV electron
equivalent (keV$_{ee}$). 
\subsection{Resolution of heat and ionisation channels}
%\label{reso}
For each detector, the baseline resolutions of the heat and the two ionisation channels are regularly controlled through runs with an automatic random trigger. These runs show that 
the noises of the three channels are not correlated. The ionisation baseline resolution can therefore be written as~:
\begin{equation}
\label{bline}
{\left(\sigma^0_{I}\right)}^2={\left(\sigma_{center}^0\right)}^2+{\left(\sigma_{guard}^0\right)}^2
\end{equation}
The $^{57}$Co calibrations give a measurement of the resolutions for the ionisation and heat signals at 122~keV. Typical values
obtained for the detectors studied here are given in Table \ref{resotab}. \\
\ \\
We parametrize the heat and ionisation signals resolutions at a given electron-equivalent energy E as~:
\begin{equation}
\label{sigt}
\sigma_{I,H}(E)=\sqrt{\left(\sigma_{I,H}^{0}\right)^2+\left(a_{I,H}E\right)^2}
\end{equation}
where the factors $a_{I}$ and $a_{H}$ are deduced from the resolution of the ionisation and heat signals at 122~keV. The resolutions of the 10.37 and 
46.52~keV peaks observed in low-background physics runs fit well with the expressions $\sigma_{I,H}(E)$ from Eq. (\ref{sigt}) 
(Fig. \ref{reso}d). It can be noted that the resolutions at $E_{I}\sim$10~keV$_{ee}$ -an energy
below which most of the WIMPs signal is expected- is dominated by the baseline
resolutions $\sigma_{I}^0$ and $\sigma_{H}^0$. \\
Finally, the recoil energy resolution can be computed from the heat and ionisation signal resolutions using Eq. (\ref{erec}). The noises of 
both signals being uncorrelated, this resolution can be written as:
\begin{equation}
\label{resrec}
\sigma_{E_{R}}=\sqrt{\left(1+\frac{V}{\varepsilon_{\gamma}}\right)^2\left(\sigma_{E_{H}}\right)^2+\left(\frac{V}{\varepsilon_{\gamma}}\right)^2\left(\sigma_{E_{I}}\right)^2}
\end{equation}
In the case of GeAl6, and for the bias voltage applied during the low-background physics run ($V$=6~V), the resolution values displayed in Table \ref{resotab} lead to 
$\sigma_{E_{R}}\sim$~8~keV FWHM around 30~keV. This value is reduced to 4~keV FWHM in the condition of the
low-background physics run recorded with GGA1 ($V$=4~V) \cite{ed2002}. 
\section{Threshold}
\label{seuil}
The ionisation and heat channel data are continuously digitized
and filtered at a rate of 200 kHz and 2kHz, respectively.
When a filtered ionisation value exceeds a fixed threshold
value, data samples in all detectors are stored to disk.
The trigger is defined by requiring a minimum threshold
on the absolute value of any of the filtered ionisation
channels.
For each event, the list of all detectors having triggered
is stored as a bit pattern.

The ionisation threshold value, $E_{I,th}$ is defined as
the ionisation energy (in keV$_{ee}$) at which the trigger
efficiency reaches 50\%.
It is the most important parameter governing the recoil energy
dependence of the efficiency. 
Its value is measured using two different techniques:
one is based on the Compton plateau observed with a $\gamma$-ray
source, and the other on coincidence neutron data.

In the first one, a $\gamma$-ray spectra is recorded using a
source producing a important Compton plateau, such as $^{60}$Co
or $^{137}$Cs.
%Monte Carlo simulations indicate that the shape of the plateau
%below 10 keV can be linearly extrapolated from higher energy.
Monte Carlo simulations indicate that the shape of the plateau
above 10 keV can be linearly extrapolated to lower energy.
The efficiency as a function of $E_I$, $\epsilon(E_I)$,
is thus obtained by dividing the measured rate by the straight
line extrapolated from the rate above 10 keV.
The resulting $\epsilon(E_I)$ data is fitted by a integral
of a gaussian (erf), yielding the experimental value of $E_{I,th}$.
However, this method is limited by the large data sample
necessary to obtain a significant number of events in the
threshold region.

\begin{figure*}
%\centerline{\includegraphics*[width=12cm,height=14cm,bb=50 157 547 707]{fig/seuilgga1.ps}
\centerline{\includegraphics*[width=12cm,height=14cm,bb=50 157 547 707]{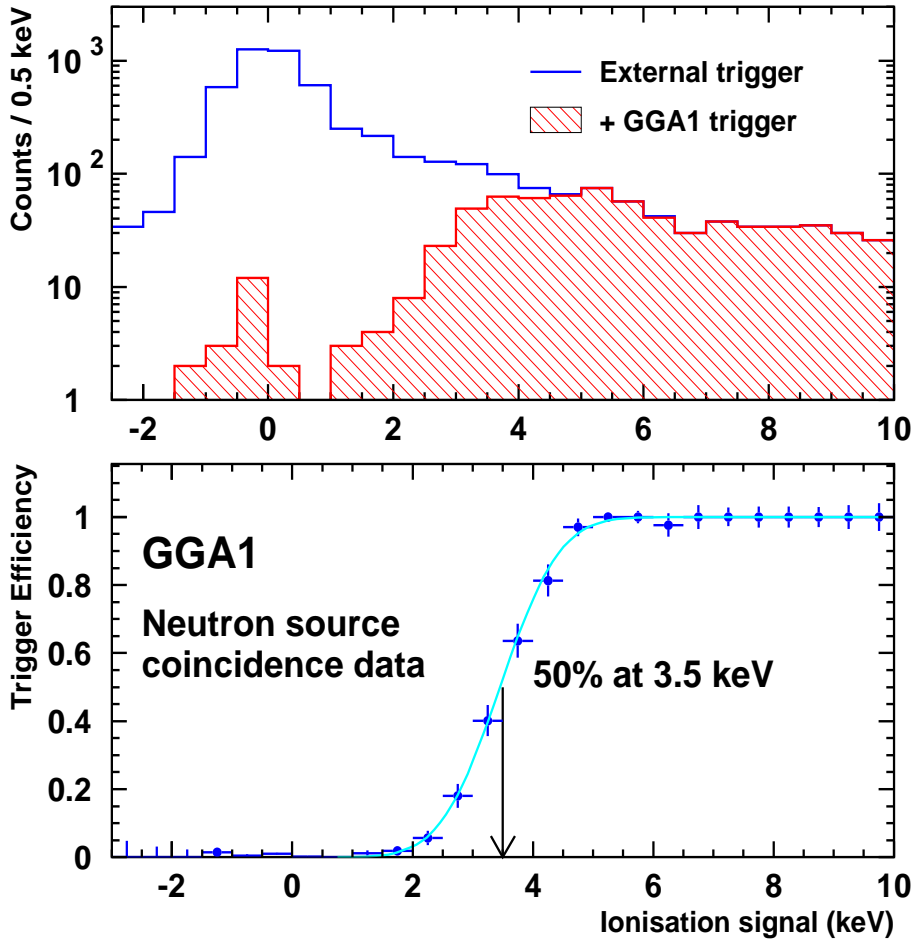}
           }
\caption{
\label{picseuil}
\textit{
         Top pannel: plot of the data recorded in the GGA1 detector during neutron calibration for any other detector (GeAl9 or GeAl10) triggering (line), and with the
         additional
         condition that GGA1 also triggers (hatched area). Bottom pannel: experimental efficiency curve for the GGA1 detector corresponding to the ratio of the 
         two distributions from the top pannel. The 100\% efficiency is reached at 5.5~keV$_{ee}$ energy,
	 corresponding to 20~keV recoil energy.
         }
        }
\end{figure*}

The second technique was made possible by the simultaneous
operation of three detectors with a $^{252}$Cf neutron source 
(and thus could not be applied to the GeAl6 detector).
Neutron scattering induces a large number of coincidence events
where at least two detectors are hit.
The upper pannel of Fig.~\ref{picseuil} shows the $E_I$ distribution
recorded in one detector with the condition that any of the other
two detector triggered (unfilled histogram).
Despite that the detector under study is not requested in the
trigger pattern, the peak at $E_I$=0 due to baseline noise
is not overwhelmingly large, due to the importance of the
coincident rate.
An unbiased sample of events with $E_I$ $>2$ keV is thus
obtained.
When in this sample it is further requested that the
detector under study be present in the trigger pattern,
the shaded histogram is obtained.
The ratio of the two distributions shown in the lower
pannel of Fig.~\ref{picseuil} correspond to the
efficiency $\epsilon(E_I)$.
This interpretation is valid in the region close
to $E_{I,th}$ and above because 
%the influence of the experimental resolution degradation 
%on the threshold value is negligeable in that energy range
% Not the point... see Jules's mail.
in that energy range the contribution of the 
peak due to baseline events is negligible 
and because the slope of the unbiased
distribution is reasonnably small compared to the
experimental resolution on $E_I$.
Indeed, applying this method to a distribution $N(E_I)$
proportionnal to $\exp(-E_I/\tau)$ and smeared with an
experimental r.m.s. resolution $\sigma$, this method would
result in a shift of $-\sigma^2/\tau$ of the deduced
value of $E_{I,th}$ relative to the true value.
In the present case, where the range of exponential slopes
and resolution are 3 $<$ $\tau$ $<$ 8 keV
and 1 $<$ 2.35$\sigma$ $<$ 2 keV, the shift should not
exceed 0.2 keV.

Both Compton and neutron coincidence techniques give consistent
ionisation threshold measurements.
The coincidence measurements are the most precise, as the neutron
source has the advantage of yielding a maximum rate
at the lowest energy, and in addition, the quenching of ionisation
for nuclear recoils ensure that the stability of the measurement
can be tested by imposing a cut on the heat signal $E_H$
without affecting the ionisation signals with $E_I$ above $\sim E_H/2$.
The measured $E_{I,th}$ values for the different ionisation
channels of the detectors under study are listed in the last column
of Table~\ref{resotab}. 
\section{Nuclear recoil band}
\label{zoneneu}
\begin{figure*}
%\centerline{\includegraphics*[width=15cm,height=15cm]{fig/rerneu.eps}
\centerline{\includegraphics*[width=15cm,height=15cm]{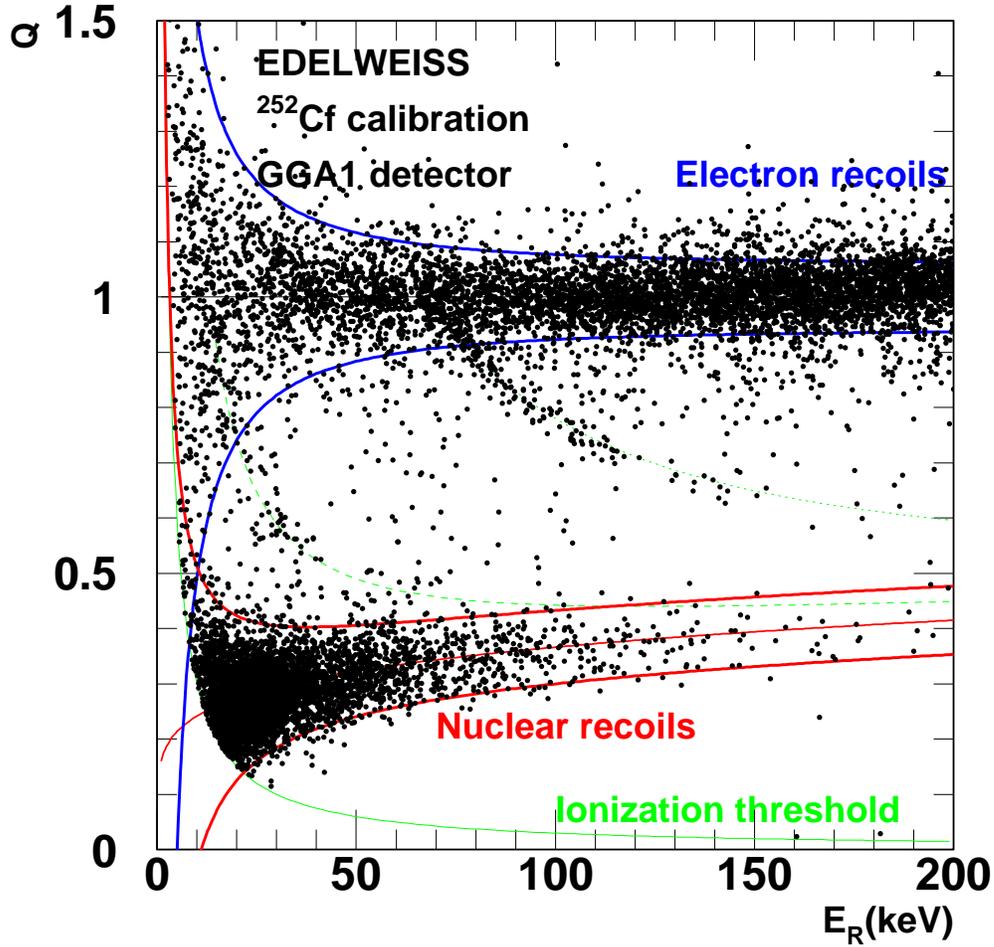}
            }
\caption{
\label{rerneu}
\textit{ 
         Projection in the ($E_R$, Q) plane of the events recorded in the GGA1 detector
         during a $^{252}$Cf calibration.
         The thick lines represent the 90\% nuclear and electronic recoils zone
         ($\pm1.645\sigma$ around $<Q_n>$ and $<Q_{\gamma}>$ respectively).
         The dotted line corresponds to the ionisation threshold curve ($E_I$=3.5~keV$_{ee}$ in this case).
         The dashed lines show where events associated with the
         inelastic scattering of neutrons on $^{73}$Ge (13.26 and 68.75~keV excited levels) are expected in
	 this plane.
        }
        }
\end{figure*}

\begin{figure*}
%\centerline{\includegraphics*[width=12cm,height=14cm,bb=50 147 547 707]{fig/agn.ps}
\centerline{\includegraphics*[width=12cm,height=14cm,bb=50 147 547 707]{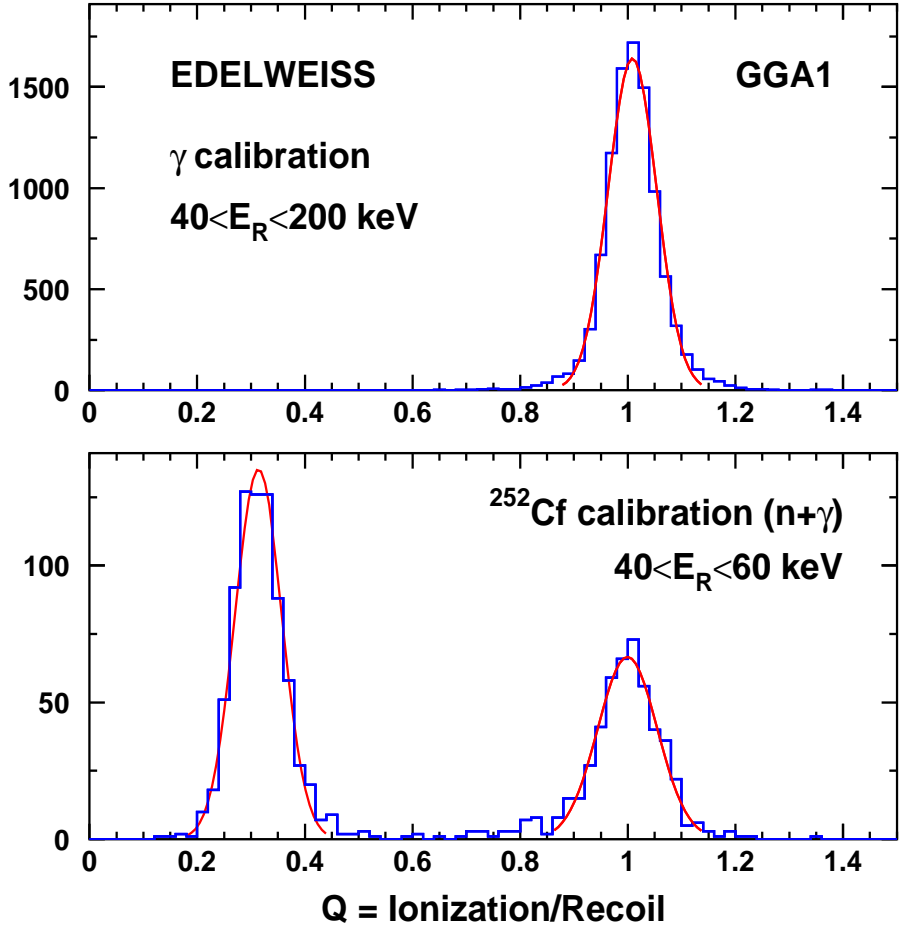}
            }
\caption{
\label{qdis}
\textit{ 
         Top pannel: spectrum of the $Q$ variable in the 40-200~keV recoil energy range 
         for events recorded in the GGA1 detector during a $^{60}$Co calibration
         (electron recoils). No events are seen for $Q<0.6$.
         This shows the excellent quality of the charge collection for this detector.
         This test 
         is performed for every detector before a low-background physics run is started.        
         Bottom pannel: Spectrum of the $Q$ variable in the 40-60~keV recoil energy range
         for events recorded in the GGA1 detector during a $^{252}$Cf 
         calibration (nuclear and electronic recoils). As the fit shows, the nuclear and
         electron recoils populations follow gaussian distributions down to 
         the 2$\sigma$ level.
        }
        }
\end{figure*}

Figure \ref{rerneu} shows a ($E_R$, $Q$) distribution from the data recorded with a 
$^{252}$Cf source emitting $\gamma$-rays and neutrons. Experimentally, the $Q$ variable appears to 
follow a gaussian distribution at the $\sim{2}\sigma$ level for both 
nuclear and electron recoils populations (Fig. \ref{qdis}). We therefore parametrize the region 
of 90\% acceptance for the nuclear recoils by the following cut:
\begin{equation}
\label{defqn}
|Q-<Q_n>|\leq{1.65\sigma_{Q_n}}
\end{equation}
where $<Q_n>$ and $\sigma_{Q_n}$ are the average value and the standard deviation of the $Q$ distribution for nuclear recoils, 
both variables being determined for each detector from $^{252}$Cf calibration data under the same experimental conditions as the
low-background physics runs.
\subsection{Neutron line}
The neutron line is the average $Q$ value for the nuclear recoils population. It is parametrized from $^{252}$Cf
calibration data by~:
\begin{equation}
\label{qneuav}
<Q_n>(E_R)=a\left(E_{R}\right)^b
\end{equation} 
The $a$ and $b$ values resulting from the fit of the experimental data for each EDELWEISS detector are statistically consistent with the values $a=0.16$ and 
$b=0.18$ quoted in \cite{heid}. The biases on the determination of $<Q_n>$ due to experimental calibration uncertainties, heat quenching 
effects \cite{sicane}, and multiple scatterings are globally taken into account with this measurement. 
\subsection{Electron and nuclear recoils zones standard deviations}
\begin{figure*}
\centerline{
            \includegraphics*[width=8cm,height=8cm]{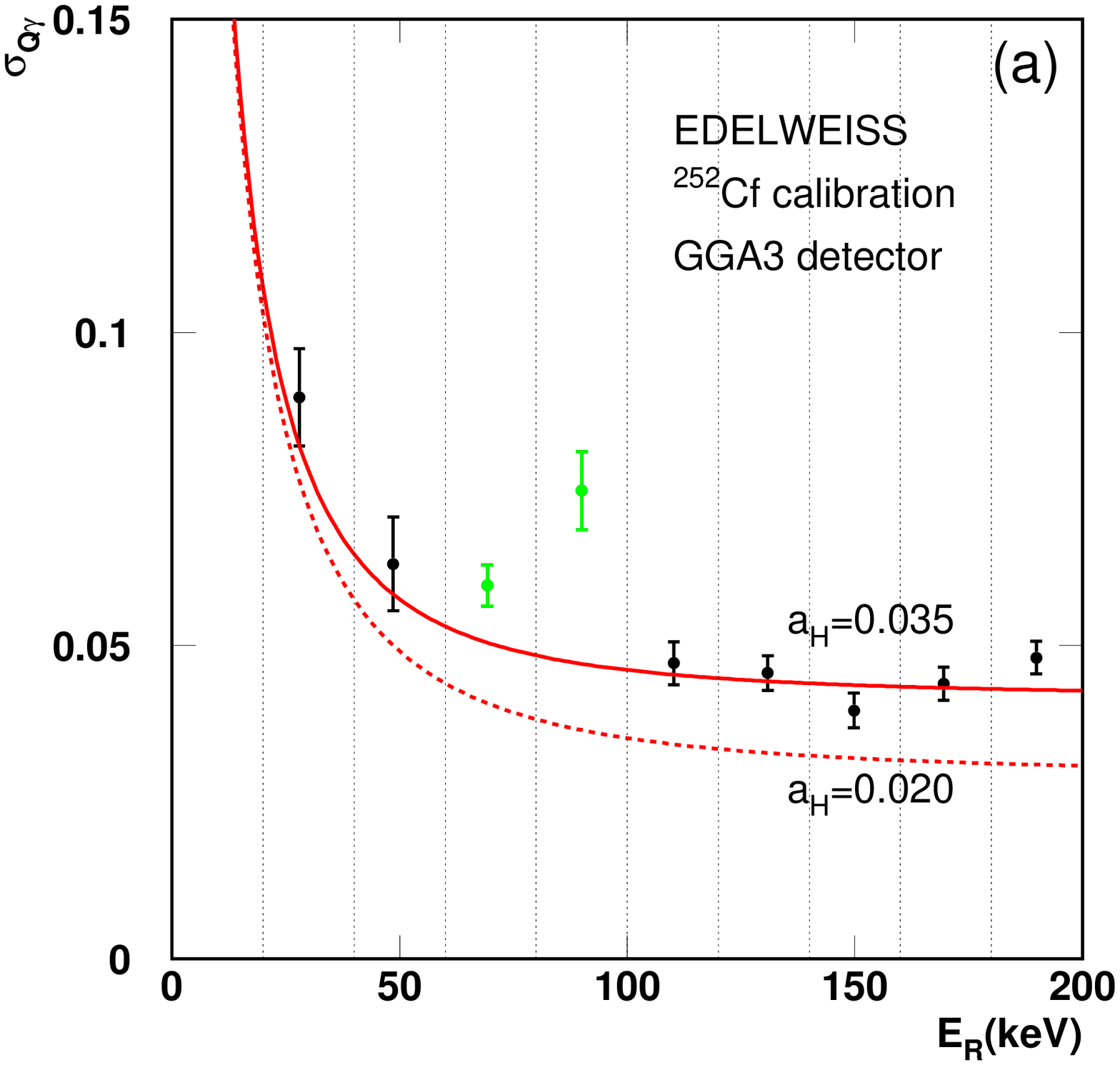} 
            \includegraphics*[width=8cm,height=8cm]{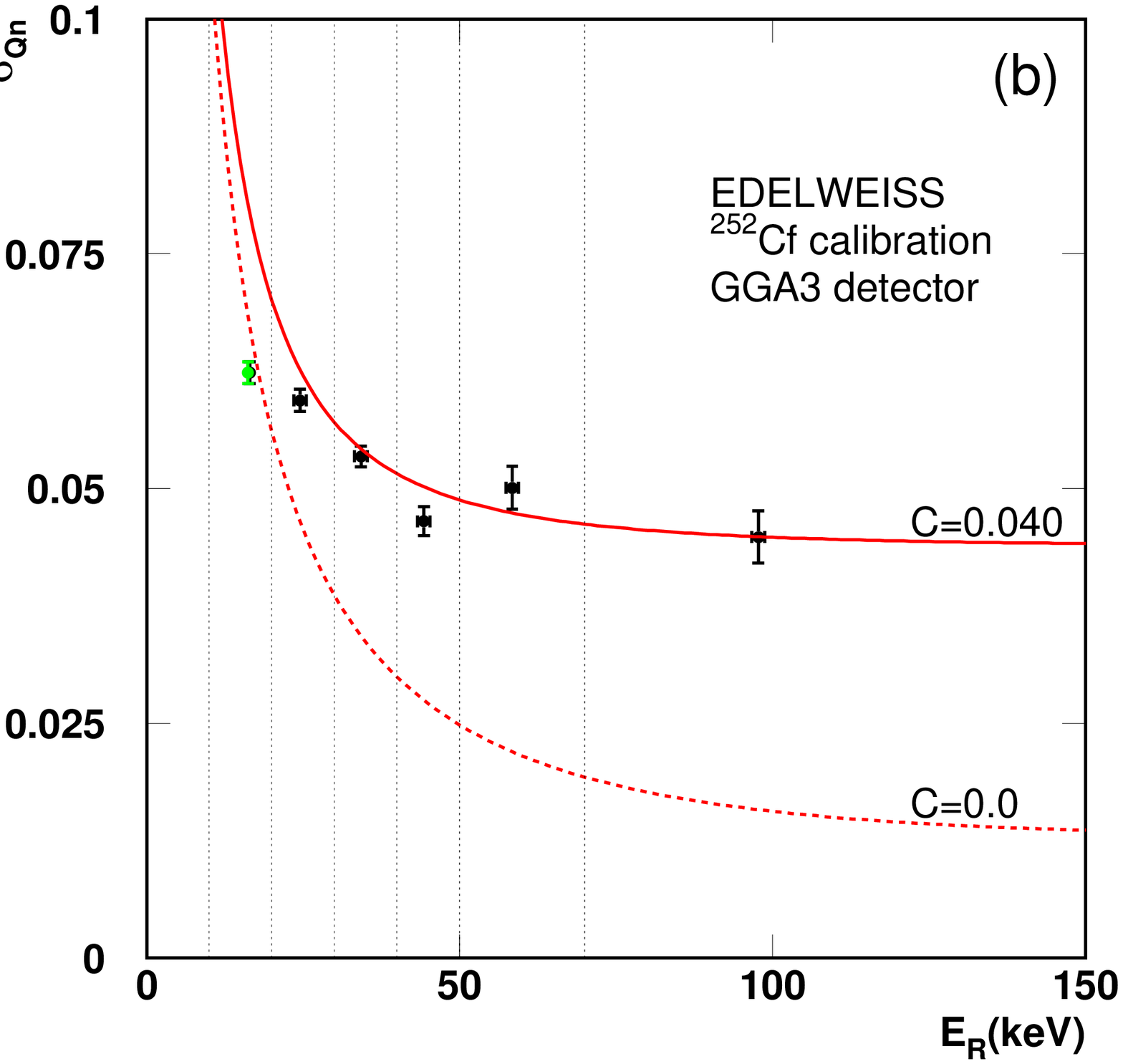}
            }
\caption{
\label{sigc}
\textit{ 
         Experimental values for $\sigma_{Q_{\gamma}}$ (a) and $\sigma_{Q_{n}}$
         (b) for a $^{252}$Cf calibration of the GGA3
	 detector. Also shown are the computed laws for these two variables from
         Eqs. (\ref{qgamma}) and (\ref{qneu1}), 
	 before (dotted line) and after (solid line) correction of the $a_H$
	 factor for (a), and for C=0 (dotted) and C=0.040 (solid) for (b). 
         The large value of $\sigma_{Q_{\gamma}}$ in (a) in the 60-80 and 80-100~keV recoil energy
	 bins (bright color) is	due to the inelastic scattering events (Fig. \ref{rerneu}),
         while the low value of $\sigma_{Q_{n}}$ in (b) in the 10-20~keV energy
         range (bright color) is related to the ionisation threshold, which artificially
         narrows the nuclear recoils distribution for low energies.
         The vertical dashed lines correspond to
	 the bins limits. 
        }
        }
\end{figure*}

The standard deviation of the electronic and nuclear recoil distributions, respectively noted $\sigma_{Q_{\gamma}}$ and $\sigma_{Q_n}$, can be 
calculated with Eqs. (\ref{erec}) and (\ref{qdef}) by propagation of the experimental values $\sigma_{I}$ and 
$\sigma_{H}$:
\begin{eqnarray}
\label{qgamma}
\sigma_{Q_{\gamma}}(E_R)&=&\frac{(1+V/3)}{E_{R}}\sqrt{{\sigma^2_{I}}+\sigma^2_{H}} \\
\label{qneu0}
\sigma^0_{Q_n}(E_R)&=&\frac{1}{{E_{R}}}\sqrt{\left(1+\frac{V}{3}<Q_n>\right)^2{\sigma^2_{I}}+\left(1+\frac{V}{3}\right)^2<Q_n>^2{\sigma^2_{H}}}
\end{eqnarray}
In the case of $^{60}$Co, $^{252}$Cf calibrations and low-background physics runs, the experimental values of $\sigma_{Q_{\gamma}}$ at high energy 
are significantly larger (up to $\sim$+30\% at 122~keV) than those calculated from the 
resolutions given in Table \ref{resotab} with Eqs. (\ref{bline}), (\ref{sigt}) and 
(\ref{qgamma}) (Fig. \ref{sigc}a). A dependance of the heat signal amplitude on the position of the
interaction provides an explanation for this discrepency. This hypothesis is consistent with the 
$\sim1$\% heat signal amplitude difference observed between center and guard events in $^{57}$Co 
calibrations. We therefore enlarge the $a_{H}$ coefficient in Eq. (\ref{sigt}) so that the analytic expression given in 
Eq. (\ref{qgamma}) for $\sigma_{Q_{\gamma}}(E_R)$ actually follows the experimental distribution for
$^{60}$Co, $^{252}$Cf and 
low-background physics runs. We have checked that 90\% of the experimental events then fall inside the 
electron recoil zone defined in this way. \\
Even after correcting the $a_H$ value, the nuclear recoils $Q$ distribution of $^{252}$Cf 
calibration data is broader at high energy than what is expected from Eq. (\ref{qneu0}) (Fig. \ref{sigc}b). 
Atomic scattering processes~\cite{lindhard}, fluctuations in the number 
of charges created by a nuclear recoil~\cite{fano} and multiple scattering (see Section \ref{multscat}) are in particular expected to 
give an intrinsic width to the $Q$ distribution for nuclear recoils and thus explain this 
behavior. The experimental $\sigma_{Q_{n}}$ dependance on recoil energy is properly described when a constant
$C$ is quadratically added to the term associated with the experimental resolution. The equation 
(\ref{qneu0}) is thus re-written as follows:
\begin{eqnarray}
\label{qneu1}
%\sigma_{Q_n}(E_R)=\frac{1}{{E_{R}}}\sqrt{\left(1+\frac{V}{3}<Q_n>\right)^2{\sigma^2_{I}}+\left(1+\frac{V}{3}\right)^2<Q_n>^2{\sigma^2_{H}}+C^2}
\sigma_{Q_n}(E_R)=\sqrt{\sigma^0_{Q_n}(E_R)+C^2}
\end{eqnarray}
Typical values of $C\sim$0.040 are determined for each EDELWEISS detector by fitting the experimental 
$\sigma_{Q_n}$ points using Eq. (\ref{qneu1}). With this definition, we have checked for each detector that 90\% 
of nuclear recoils induced by $^{252}$Cf calibrations are inside the nuclear recoil zone defined in Eq.
(\ref{defqn}). 
\subsection{Effect of multiple scattering}
\label{multscat}
\begin{figure*}
\begin{center}
\begin{tabular}[h]
{c}
            \includegraphics*[width=10cm,height=6.5cm, bb=130 400 507 647]{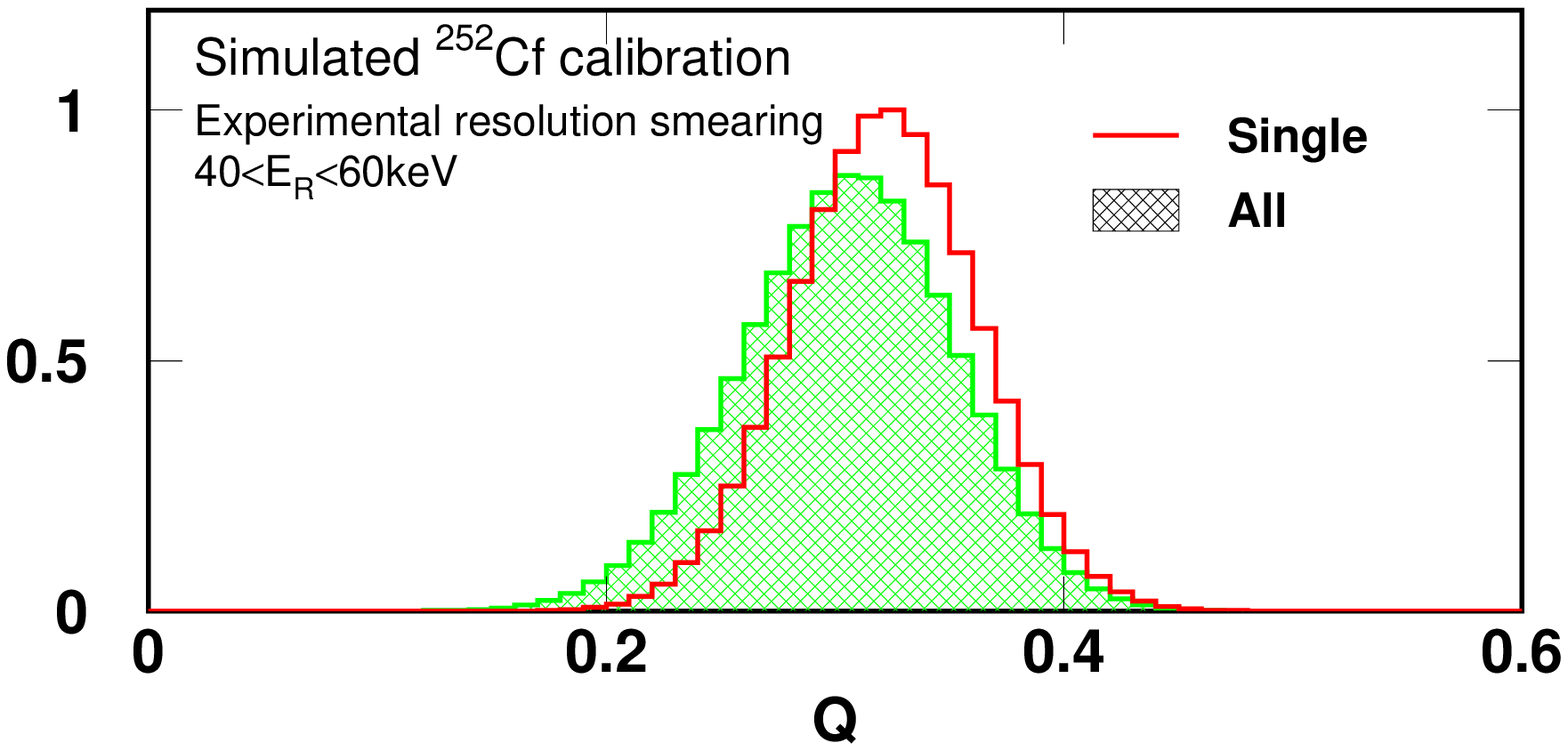} \\
            \includegraphics*[width=10cm,height=6.5cm, bb=130 420 507 647]{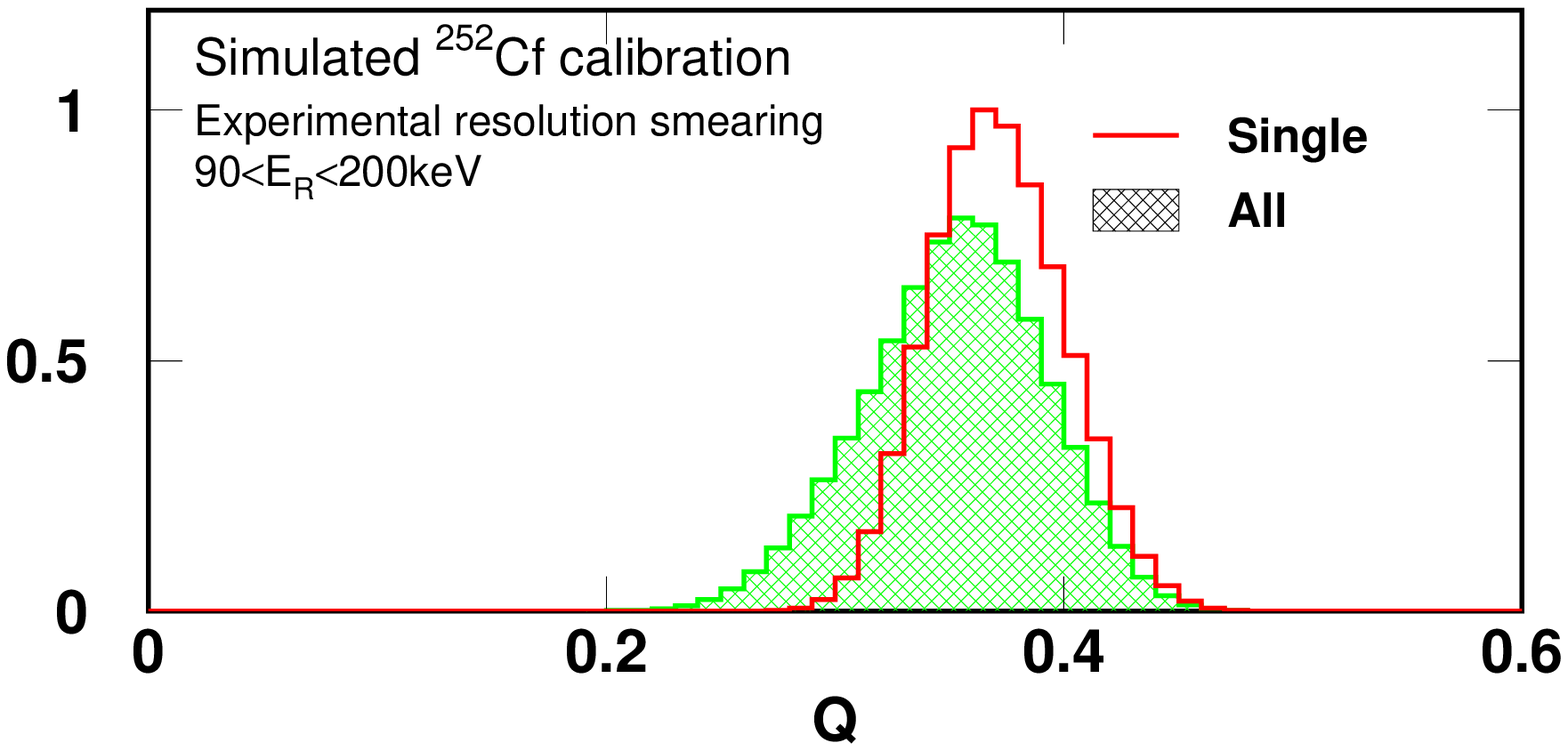}
 
\end{tabular}	    
\end{center}
\caption{
\label{simurer}
\textit{ 
         Normalized spectra of the Q variable for a GEANT simulation of a $^{252}$Cf calibration, selecting
	 (hatched area) or not (continuous line) single interactions, in the 40-60~keV (top pannel) and 90-200~keV 
	 (bottom pannel) energy ranges. The distribution corresponding to all
	 interactions is slightly shifted down ($\sim$~0.015 units between 20 and 200 keV) in regards to the 
	 single interactions distribution, and is only slightly broader. 
        }
        }
\end{figure*}

The nuclear recoil zone is determined through neutron calibrations, for which the proportion of multiple
interactions is around 40\% between 20 and 200~keV. This is of particular importance because in
contrast to neutrons, WIMPs are expected to interact only once in the detector, and the $Q$ variable 
is in this case larger than when the same energy is deposited 
in multiple nuclear interactions, as can be deduced from Eq. (\ref{qneuav}). \\
We therefore evaluated quantitatively the effect of multiple interactions using a GEANT~\cite{geant} 
simulation of $^{252}$Cf calibrations of the EDELWEISS detectors. The $Q$ variable has been calculated for the simulated nuclear events by associating with 
Eq. (\ref{qneuav}) an ionisation signal of amplitude $e_{I}=0.16\left(e_{R}\right)^{1.18}$ to an energy deposit $e_{R}$ in a single interaction, and summing each 
individual $e_I$ to obtain the total $E_I$ energy for a given neutron. 
The effect of multiple interactions has then been evaluated with these simulated data by smearing the
resulting $Q$ distribution with the experimental resolution given in Eq. (\ref{qneu0}), and then by 
comparing the distributions obtained when selecting or not single interactions events (Fig. \ref{simurer}). \\
Although multiple interactions tend to lower $<Q_n>$, this effect remains weak, and the $Q$ distribution 
associated with single interactions events is only slightly narrower and completely included in the wider band. The nuclear recoils
zone determined through $^{252}$Cf calibrations has therefore been conservatively used for the low-bakground 
physics run analysis. 
\subsection{Analysis energy range}
Equations (\ref{qgamma}) and (\ref{qneu0}) predict that the discrimination between electronic and nuclear
interactions is deteriorated at low energies (see also Fig. \ref{rerneu}). Rejection of the $\gamma$-ray
background at a given level therefore defines a lower bound for the analysis energy range. \\
Secondly, the detection efficiency has to be as close to 100\% as possible in the analysis window in order to insure a good quality 
for the data set. The trigger threshold is therefore another factor which has to be taken into account for the definition of the 
analysis lower energy bound. For both 2000~\cite{ed2000} and 2002~\cite{ed2002} runs, the choice of the analysis 
lower bound has mainly been driven by this last factor. The threshold values of 5.7 
and 3.5~keV$_{ee}$ for the ionisation signal indeed correspond respectively to recoil energies of 30 and 20~keV for a 100\% detection 
efficiency, and 90\% efficiency when the nuclear recoil zone is taken into account. \\
Extensive $\gamma$-rays and neutron calibrations are performed before the physics data taking is initiated 
in order to fix the lowest recoil energy value corresponding to acceptable levels of $\gamma$-ray background 
rejection and detection efficiency. This ensures that the lower limit of the analysis 
window is not influenced by the possible presence of events in the final data set. The definition of the upper
bound of the analysis window is described in \cite{ed2002}.
\section{Fiducial volume}
\label{volfid}
\subsection{Modelisation of the collection process}
\label{deffid}
The segmentation of the upper charge collection electrode in a central part and a guard ring leads to the 
definition of a fiducial volume. This volume is shielded against a significant amount of the radioactivity of the detector environment by the
peripherical volume, as shown in \cite{ed2000}. To allow for the experimental resolution on the ionisation signals, the fiducial cut is defined as 
corresponding to a fraction of 3/4 of the charge collected on the center electrode. In order to give a robust and precise 
estimation of the detector volume
associated with this fiducial cut, it is necessary to relate a given ratio of the two ionisation signal amplitudes to a given volume 
inside the detector. This is not a straightforward process: first, for non-WIMP interactions, multiple
interactions have to be taken into account, and furthermore, interactions between charges may play a crucial 
role in the collection process. In particular, the important proportion of events with a charge signal shared between the two channels observed in each
detector for $^{60}$Co calibrations (see e.g. Fig. 3) hints to the importance of these charge interactions processes. \\
In order to test their influence on the determination of the fiducial volume, we choose to model the
collection process with the simplified phenomenological description of charge collection given in Ref. \cite{penn}, associated with the hypothesis of a
plasma effect before charge drift. We will see that, even if some of our results cannot
be explained in the framework of this very simplified model (Section \ref{resfid}), it provides a good empirical tool to determine the fiducial volume and estimate 
systematic errors on its value (Section \ref{valfid}). The model used here assumes the distribution of the charges in a sphere with uniform density, extending to a maximal 
radius $r_b$ before the charge is fully collected. Charges are distributed among the two electrodes depending on the position of the interaction relative to the surface
corresponding to the separation between drift lines going to the center and guard rings. Here, we assume for
simplicity that this surface is parametrized by a
cylinder of radius $R_C$ (Fig. \ref{vfid}). For an interaction at the radius $R>R_C+r_b$ in the crystal, the whole charge is fully associated with the guard ring. 
If $R<R_C-r_b$, then the charge has to be associated with the center electrode. Finally, if $R_C+r_b>R>R_C-r_b$, then the charge is splitted among the 
two electrodes, with a relative proportion associated with the center electrode corresponding to the fraction of the sphere inside the cylinder of radius $R_C$. 
For given values of $r_b$ and $R_C$, the fiducial volume is determined in this model by the following
expression of the fiducial radius:
\begin{equation}
\label{rfid}
R_{fid}=R_C+2\cos\left({\frac{13\pi}{9}}\right)r_b
\end{equation}
A fraction of 1/4 of the total volume of a sphere of radius $r_b$ centered on $R_{fid}$ is inside the 
cylinder of radius $R_{C}$. In the framework of our model, interactions inside the cylinder of radius $R_{fid}$ 
thus correspond to a charge collection equal or greater than 3/4 of the total charge.
\subsection{Validity and limits of the modelisation}
\label{resfid}
In order to test its ability to reproduce the distribution of charge amplitudes, 
ionisation signals are simulated in the framework of this simple model, using the program GEANT~\cite{geant} for $^{60}$Co and $^{252}$Cf calibrations, 
as described in Section \ref{multscat}. The
parameters $r_b$ and $R_C$ of the simulated data are then adjusted to match the experimental distribution of
the $Y$ variable on a given energy range, the $Y$ variable being defined 
as the normalized difference of the ionisation signals: 
\begin{equation}
Y=\frac{E_{guard}-E_{center}}{E_{guard}+E_{center}}
\end{equation}
The result of this optimisation is shown in Fig. \ref{super} in the case of a $^{60}$Co calibration of the GeAl6 detector under
6.3~V bias voltage. The shape of the simulated distribution closely follows that of the experimental data, while a simulation 
using an alternative model (linear distribution of the charge, detailed in \cite{ltd9}) clearly exhibits a different pattern. \\
\ \\
We have also studied the evolution of the ($r_b$, $R_C$) parameters as a function of bias voltage for the GeAl6 detector~\cite{these}.
$R_C$ should not depend on the value and sign of the bias voltage, since it is related to the static field distribution only, 
while $r_b$ should increase with decreasing bias
voltage: as the field increases, the less time there is for diffusion processes. The values of the parameters $r_b$ and $R_C$ determined for 
$^{60}$Co and $^{252}$Cf calibrations of the GeAl6 detector versus the applied
field are displayed in Fig. \ref{rbrc}. The $r_b$ and $R_C$ values follow the expected behavior. Moreover, the 
mean measured value of $R_C$ ($<R_C>=24.45\pm0.05$~mm) is statistically compatible with the value $R_{electro}=24.4$~mm expected from numerical calculations 
of the electric potential inside this detector (see Table \ref{det}). \\
The very large $r_b$ values are a clear sign that macroscopic charge extension perpendicular to the drift direction occurs before the charge collection is completed.
However the data does not support that this expansion is driven by the plasma effect invoqued in Ref. \cite{penn}: a charge cloud size of the order or above a
millimeter is indeed not compatible with results of studies on the dead layer [5, 16]. Furthermore, Fig. \ref{rbrc} shows that $r_b(-)<<r_b(+)$ and 
that the values of $r_b$ for $^{60}$Co and $^{252}$Cf calibrations do not differ significantly for a same bias voltage. These two experimental results are also in strong
disagreement with the predictions derived from the hypothesis of a plasma effect: firstly, the observed asymmetry for $r_b$ values 
between positive and negative bias voltage does not find any explanation in the
framework of the plasma model, and secondly, the plasma effect should be weaker in the case of $^{60}$Co 
calibrations than for $^{252}$Cf (and thus $r_b$ values much smaller), since 
$\gamma$-rays induce much lower charge densities than neutrons. 
These are strong indications that the simple model presented here does not provide a proper description 
of the dynamics of the charge drift and collection. Charge repulsion during drift, not taken into account
here, could for example play an important role in the collection process. A more detailed study, with
dedicated detectors, has been initiated in the EDELWEISS collaboration in the aim of better understanding 
the collection process~\cite{alex}. 
\subsection{Measurement of the fiducial volume}
\label{valfid}
\begin{figure*}
%\centerline{\includegraphics*[width=15cm,height=15cm]{fig/super_n.eps}
\centerline{\includegraphics*[width=15cm,height=15cm]{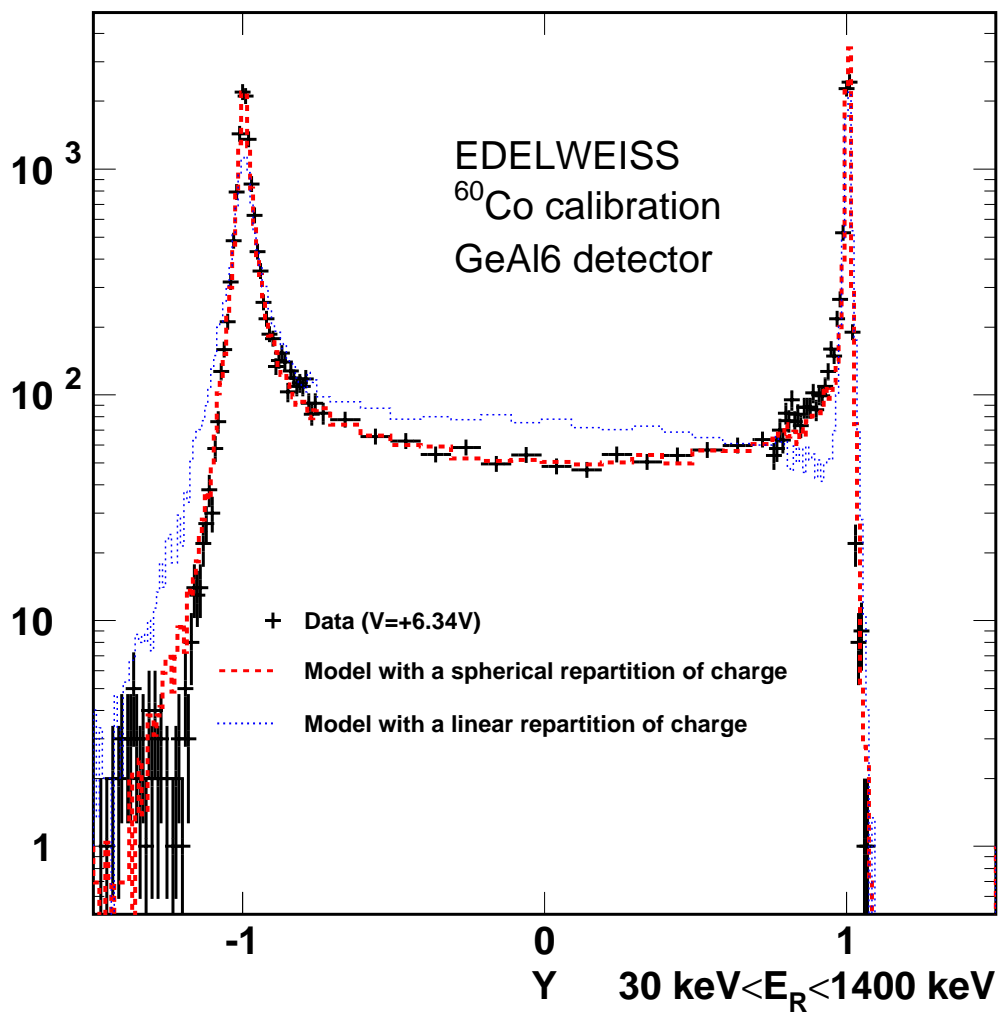}
            }
\caption{
\label{super}
\textit{ 
         Distribution of the $Y=\frac{E_{guard}-E_{center}}{E_{guard}+E_{center}}$ variable for events of $^{60}$Co calibration under
	 +6,3~V bias voltage with 30~keV$<E_I<$1400~keV (cross). The 
         two other spectra correspond to the simulated distributions obtained for modelisations in the case of a charge reparted in a sphere (dashed, $r_b$=4.2~mm, 
         $R_C$=24.5~mm) and charge reparted linearly (dotted, for parameters giving the same fiducial volume).
        }
        }
\end{figure*}

\begin{figure*}
%\centerline{\includegraphics*[width=15cm,height=15cm]{fig/rbrc.eps}
\centerline{\includegraphics*[width=15cm,height=15cm]{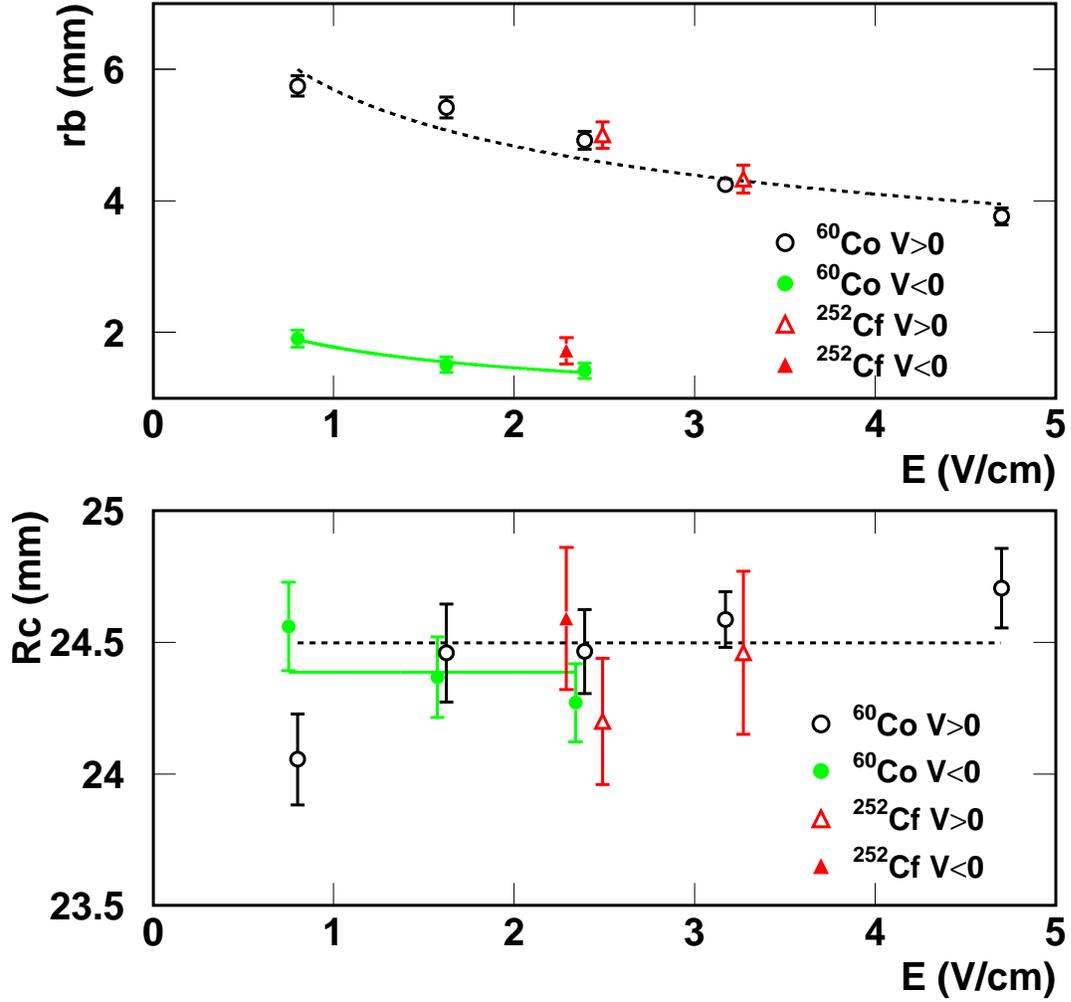}
            }
\caption{
\label{rbrc}
\textit{ 
         Optimized values for $r_b$ (top pannel) and $R_C$ (lower pannel) for $^{60}$Co (circles) and $^{252}$Cf (triangles) calibrations of the GeAl6 
         detector versus applied field values. Positive (negative) bias voltages correspond to the empty (filled) symbols. The $r_b$ distribution 
         is fitted by a power law: $r_b=aE^{-b}$. The fit gives $a_+=5.7\pm0.1$, $a_-=1.8\pm0.1$, $b^+=-0.24\pm0.02$ and $b^-=-0.28\pm0.09$. 
        }
        }
\end{figure*}

\begin{figure*}
%\centerline{\includegraphics*[width=15cm,height=15cm]{fig/vfidd.eps}
\centerline{\includegraphics*[width=15cm,height=15cm]{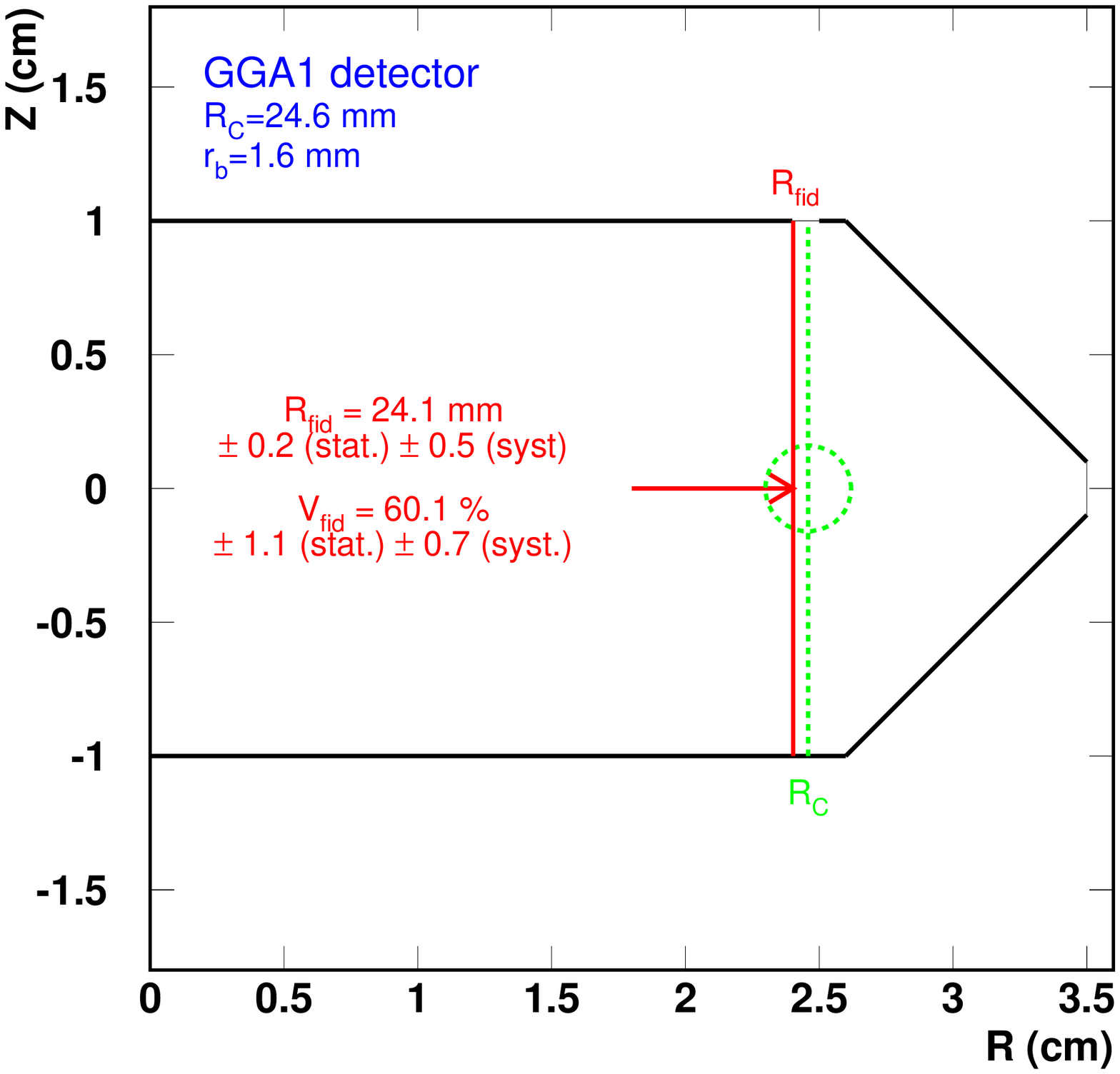}
            }
\caption{
\label{vfid}
\textit{ 
         Representation of the GGA1 detector in the (R,Z) plane. The thick line corresponds to
	 $R_{fid}=24.1~mm$. Also shown is the line $R_C=24.6$~mm delimiting the volumes associated with
	 collection on the center and guard electrodes 
         (dotted) and a circle of radius $r_b$=1.6~mm centered on $R_C$ (dotted).  These values correspond to the optimum for a $^{252}$Cf calibration of
	 the GGA1 detector under a bias voltage of -4.00~V.
        }
        }
\end{figure*}

Our results clearly point out the limits of the modelisation presented in Section \ref{deffid}. Still, it has to be stressed that this model reproduces 
correctly the distribution of charges 
among the electrodes (Fig. \ref{super}), which represents the net effect of the charge collection process. It is therefore 
sufficient to give a precise determination of the fiducial volume and evaluate possible systematic errors, before a better, physically motivated model replaces 
it. \\
We have calculated $R_{fid}$ with Eq. (\ref{rfid}) and the $r_b$ and $R_C$ values determined
from $^{252}$Cf calibrations under the same bias voltage as that of the low-background physics run for each detector. 
For all detectors except GeAl10, the $R_C$ values are compatible with those expected from the geometry of 
the electrodes and from numerical simulations of the electric potential inside the detectors. The 
values determined for $R_{fid}$ for the detectors are summarized in Table \ref{vfidt}. The systematic error associated with the uncertainty on the exact
mechanism producing the charge expansion is evaluated by taking the difference between the fiducial volume 
value deduced using the linear model and the one presented in Section \ref{deffid}. Despite
the poor description of the charge distribution by the linear model (see Fig. \ref{super}), this difference is only 1\%. The variation of 
the energy range used to determine the values of $r_b$ and $R_C$ through comparison of the experimental and simulated 
$Y$ distributions proved to be a minor contribution to this systematic error. \\
\ \\
An alternative evaluation of the fiducial volume is
the fraction of cosmic activation events at 8.98 and 10.37~keV
(see Fig.~\ref{reso}) selected by the fiducial volume cut. 
Such events are expected to be evenly spread inside the detector,
and are observed at rates varying between 3 to 15 events per
detector per day.
In the few days following a neutron calibration, the 10.37~keV rate is
also enhanced due to $^{71}Ge$ activation ($T_{1/2}$ = 2.7 d),
a population that is also expected to be evenly spread inside the
detector.
The measured fractions, directly interpreted as $V_{fid}$ values,
are listed in the last column of Table~\ref{vfidt}.
They are compatible within statistics with the values derived from
the neutron calibration data and the collection process
modelisation. The cosmic activation data is however less precise due to
statistics, but this measurement is a good cross-check for the determination of the fiducial volume, and validates the use of the model presented in Section
\ref{deffid} to determine the fiducial volume.
\section{Conclusion}
We have described in the present work the calibration aspects of the data analysis in the EDELWEISS  
experiment. In particular, the nuclear recoil zone and fiducial volume have been estimated using several
methods, allowing to define a conservative value of these important parameters. A simple parametrization allows us to reproduce accurately the
distribution of the charges between the centre and guard electrodes associated with $^{60}$Co 
and $^{252}$Cf calibrations, making possible the systematic studies necessary to establish the robustness of the determination of the fiducial volume of the
detectors.
%
%
%
%
%%%%%%%%%%%%%%%%%%%%%%%%%%%%%%%%%%%%%%%%%%%%%%%%%%%%%%%%%%%  BIBLIOGRAPHIE  %%%%%%%%%%%%%%%%%%%%%%%%%%%%%%%%%%%%%%%%%%%%%%%%%%%%%%%%%%%%%%%%%%%
%
%

%
\newpage
%
%
%%%%%%%%%%%%%%%%%%%%%%%%%%%%%%%%%%%%%%%%%%%%%%%%%%%%%%%%  TABLES  %%%%%%%%%%%%%%%%%%%%%%%%%%%%%%%%%%%%%%%%%%%%%%%%%%%%%%%%%%%%%%%%%%%%%%%%%%%%%%%
%
%
\begin{table}[h]
\begin{center}
\begin{tabular}[h]
{|c|c|c|c|c|c|}
\hline
Label&Mass&$R_{electro}$&Vol. NTD&Amorphous&$T_{running}$ \\
&(g)&(mm)&(mm$^3$)&layer&(mK) \\
\hline
GeAl6&321.62&24.4&4.0&none&27 \\
\hline
GeAl9&325.43&24.0&5.6&none&17 \\
\hline
GeAl10&323.91&24.0&5.6&none&17 \\
\hline
GGA1&318.50&24.0&1.64&Ge&17 \\
\hline
GGA3&324.40&24.0&5.6&Ge&17 \\
\hline
GSA1&313.68&24.0&5.6&Si&17 \\
\hline
GSA3&297.03&24.0&5.6&Si&17 \\
\hline
\end{tabular}
\end{center}
\caption{
         \label{det}
         \textit{ 
                 Main parameters for the EDELWEISS detectors studied in this article. "$R_{electro}$" refers
		 to the radius value of the cylindrical volume associated with charge collection
		 on the center electrode.
		 These values are calculated through electrostatic simulation of the detector, taking into
		 account the actual electrodes geometry. The existence 
		 of an amorphous Ge or Si layer under the electrodes is also mentionned. "$T_{running}$" is 
		 the value of the regulated cryostat temperature while running. 
		 }
         }    
\end{table}
\ \\
\ \\
\begin{table}[h]
\begin{center}
\begin{tabular}[h]
{|c||c|c|c||c|c||c|c|}
\hline
&\multicolumn{3}{c||}{FWHM @ 0~keV}&\multicolumn{2}{c||}{FWHM @ 122~keV}&\multicolumn{2}{c|}{Trigger Threshold} \\
\hline
&Center&Guard&Heat&Ion.&Heat&Center&Guard \\
Detector&(keV$_{ee}$)&(keV$_{ee}$)&(keV$_{ee}$)&(keV$_{ee}$)&(keV$_{ee}$)&(keV$_{ee}$)&(keV$_{ee}$) \\
\hline
GeAl6&2.0&1.4&2.2&2.8&3.5&6.0&4.0 \\
\hline
GeAl9&1.2&1.4&0.5&2.6&3.3&4.3&4.9 \\
\hline
GeAl10&1.1&1.3&0.4&3.0&3.5&3.3&4.3 \\
\hline
GGA1&1.3&1.3&1.3&2.8&3.5&3.5&3.5 \\
\hline
GGA3&1.3&1.5&0.4&3.1&2.7&2.9&3.9 \\
\hline
GSA1&1.2&1.4&0.6&3.1&2.8&3.5&3.4 \\
\hline
GSA3&1.1&1.3&1.4&3.3&3.3&3.0&3.4 \\
\hline
\end{tabular}
\end{center}
\caption{
         \label{resotab}
         \textit{ 
                 Typical values obtained in keV$_{ee}$ for the full width half maximum resolution for heat and ionisation signals at 0 and 122~keV 
		 for the detectors studied in this article. The precision on these measurements are $\pm0.1$~keV at 0~keV and 
		 $\sim\pm$0.2~keV at 122~keV. 
		 Also given here are the threshold values for the center and guard channels. The precision is $\pm$0.1 keV for both channels,
		 except for GeAl6 where it is $\pm$0.5 keV. 
                 }
         }    
\end{table}
\ \\
\ \\
\begin{table}[h]
\begin{center}
\begin{tabular}[h]
{|c||c|c|c|c|c|c|}
\hline
Detector&Bias&$r_b$&$R_C$&$R_{fid}$&$V_{fid}$&Activation \\
&(V)&(mm)&(mm)&(mm)&(\%)& $V_{fid}$(\%) \\
\hline
GeAl6&+6.34&$4.3\pm0.2$&$24.5\pm0.3$&$23.0\pm0.3$&$54.6\pm1.4$&$50\pm3$ \\
\hline
GeAl9&+2.00&$6.1\pm0.2$&$23.5\pm0.4$&$21.4\pm0.4$&$47.4\pm2.0$& $53\pm4$ \\
\hline
GeAl10&-3.00&$2.3\pm0.2$&$21.9\pm0.4$&$21.1\pm0.4$&$46.0\pm1.7$&$50\pm4$ \\
\hline
GGA1&-4.00&$1.6\pm0.1$&$24.6\pm0.2$&$24.1\pm0.2$&$60.1\pm1.1$&$57\pm3$ \\
\hline
%GGA3&-4.00&$1.5\pm0.1$&$24.1\pm0.2$&$23.6\pm0.2$&$57.5\pm1.2$& $63\pm3$ \\
 GGA3&-4.00&$1.4\pm0.1$&$24.1\pm0.2$&$23.6\pm0.2$&$57.7\pm0.7$& $60\pm5$ \\
\hline
%GSA1&-4.00&$1.7\pm0.3$&$24.1\pm0.5$&$23.5\pm0.5$&$57.2\pm2.6$& $58\pm6$\\
 GSA1&-4.00&$1.5\pm0.1$&$24.2\pm0.2$&$23.7\pm0.2$&$58.3\pm0.8$& $61\pm4$\\
\hline
%GSA3&-4.00&$1.9\pm0.3$&$24.0\pm0.6$&$23.3\pm0.6$&$56.4\pm2.8$& $62\pm6$ \\
 GSA3&-4.00&$1.5\pm0.1$&$23.9\pm0.2$&$23.3\pm0.2$&$56.2\pm0.8$& $61\pm5$ \\
\hline
\end{tabular}
\end{center}
\caption{
         \label{vfidt}
         \textit{ 
                 Values of various parameters for the EDELWEISS bolometers determined from $^{252}$Cf calibrations under the given bias voltage. The 
                 error bars correspond to statistical errors. The systematic error on $V_{fid}$ is $\sim$1\%.
		 "Activation" refers to the fraction of 8.98 and 10.34~keV events recorded with the
		 fiducial volume cuts ($E_{center}>3E_{guard}$).
                 }
         }    
\end{table}
\newpage
%
%%%%%%%%%%%%%%%%%%%%%%%%%%%%%%%%%%%%%%%%%%%%%%%%%%%%%%%%  FIGURES  %%%%%%%%%%%%%%%%%%%%%%%%%%%%%%%%%%%%%%%%%%%%%%%%%%%%%%%%%%%%%%%%%%%%%%%%%%%%%%%
%
%
%
% The dotted lines correspond to total ionisation signals $E_I$=1171~keV and $E_I$=1332~keV. 
%

\begin{thebibliography} {99}
\bibitem{berg}
L. Bergstr\"om, Rep. Prog. Phys {\bf 63}, 793 (2000).
\bibitem{xfn} 
X.F.~Navick {\it et al.}, NIM A {\bf 444}, 361 (2000).
\bibitem{ed2000}
A. Beno\^{\i}t {\it et al.}, Phys. Lett. B {\bf 479} 8 (2000).
\bibitem{ed2002}
A. Beno\^{\i}t {\it et al.}, Phys. Lett. B  {\bf 545} 43 (2002).
\bibitem{surf} 
P. Luke {\it et al.}, IEEE Trans. Nucl. Sci. 41 (4) (1994) 1074. \\
T. Shutt {\it et al.}, NIM A {\bf 444}, 340 (2000). \\
T. Shutt {\it et al.}, in Proc. {\it 9$^{th}$ Int. Workshop on Low Temperature Detectors}, AIP conference proceedings {\bf 605}, 513 (2001).
\bibitem{N03} 
XF. Navick {\it et al.}, to be published.
\bibitem{lneg}
B. Neganov and V.~Trofimov, USSR patent No 1037771, Otkrytia i izobreteniya {\bf 146}, 215 (1985). \\
P.N.~Luke, J. Appl. Phys. {\bf 64}, 6858 (1988).
\bibitem{heid}
P. Di Stefano {\it et al.},  Astropart. Phys. {\bf 14}, 329 (2001). 
\bibitem{sicane}
E.~Simon {\it et al.}, NIM A {\bf 507}, 643 (2003).
\bibitem{geant}
R.~Brun {\it et al.}, {\it GEANT3}, CERN report DD/EE/84-1 (1987).
\bibitem{lindhard}
J.~Lindhard {\it et al.}, Mat. Phys. Medd. Dan. Vid. Selsk {\bf 10}, 1 (1963). 
\bibitem{fano}
T.~Yamaya {\it et al.}, NIM {\bf 159}, 181 (1979).
\bibitem{penn}
M.J. Penn {\it et al.}, in Proc. {\it 6$^{th}$ Int. Workshop on Low Temperature Detectors}, NIM A {\bf 370}, 215 (1996).
\bibitem{ltd9}
O.~Martineau {\it et al.}, in Proc. {\it 9$^{th}$ Int. Workshop on Low Temperature Detectors}, AIP conference proceedings {\bf 605}, 505 (2001).
\bibitem{these}
O.~Martineau,~Recherche~de~WIMPs~par~l'exp\'erience EDELWEISS: caract\'erisation~des~d\'etecteurs~et~analyse~des donn\'ees,
PhD thesis, Universit\'e Lyon I (2002) (in french). \\ Available at~http://edelweiss.in2p3.fr/pub/fichiers/theses.html.
\bibitem{alex}
A.~Broniatowski {\it et al.}, in Proc. {\it 9$^{th}$ Int. Workshop on Low Temperature Detectors}, AIP conference proceedings {\bf 605}, 521 (2001). \\
A.~Broniatowski, to be published in LTD10 proceedings (Genova, July 2003).
%
\end{thebibliography}
\end{document}